\documentclass[runningheads,a4paper]{llncs}

\usepackage{amssymb}
\usepackage{adjustbox}
\usepackage{graphicx}
\usepackage{siunitx}
\usepackage{mathtools}
\usepackage{threeparttable}
\usepackage{float}
\usepackage{booktabs}
\usepackage{tabularx}
\usepackage{makecell}
\newcolumntype{C}[1]{>{\centering\arraybackslash}p{#1}}
\newcolumntype{L}[1]{>{\raggedright\arraybackslash}p{#1}}
\usepackage{multicol}

\usepackage{url}
\newcommand{\keywords}[1]{\par\addvspace\baselineskip
\noindent\keywordname\enspace\ignorespaces#1}
\usepackage{subcaption}
\graphicspath{ 
{arch/} 
{fig_boxPlots/} 
{fig_boxPlots/unet_attention/} 
{fig_boxPlots/unet_25D/} 
{fig_boxPlots/lstm_unet/} 
{fig_boxPlots/lstm_resnet34/} 
{fig_boxPlots/lstm_cornet/} 
{fig_images/}
{fig_motionSimPipeline/}
{fig_images/BraTS19_CBICA_AQG_1/} 
{fig_images/BraTS19_CBICA_AYU_1/} 
{fig_images/BraTS19_TCIA02_309_1/} 
{fig_images/BraTS19_TCIA08_319_1/}
{fig_trainValSummary/}
{data_split/}
}

\begin{document}

\mainmatter

\title{Assessing Lesion Segmentation Bias of Neural Networks on Motion Corrupted Brain MRI}

\titlerunning{Assessing Lesion Segmentation Bias on Motion Corrupted Brain MRI}

\author{Tejas Sudharshan Mathai\inst{1,3}, Yi Wang\inst{1,2,3}, Nathan Cross\inst{2}, }

\authorrunning{T.S. Mathai, Y. Wang et al.}

\institute{Philips Healthcare, Bothell WA 98201, USA \and Department of Radiology, University of Washington, Seattle WA 98195 \and Equal contribution}

\maketitle

\begin{abstract}

Patient motion during the magnetic resonance imaging (MRI) acquisition process results in motion artifacts, which limits the ability of radiologists to provide a quantitative assessment of a condition visualized. Often times, radiologists either ``see through" the artifacts with reduced diagnostic confidence, or the MR scans are rejected and patients are asked to be recalled and re-scanned. Presently, there are many published approaches that focus on MRI artifact detection and correction. However, the key question of the bias exhibited by these algorithms on motion corrupted MRI images is still unanswered. In this paper, we seek to quantify the bias in terms of the impact that different levels of motion artifacts have on the performance of neural networks engaged in a lesion segmentation task. Additionally, we explore the effect of a different learning strategy, curriculum learning, on the segmentation performance. Our results suggest that a network trained using curriculum learning is effective at compensating for different levels of motion artifacts, and improved the segmentation performance by $\sim$9\%-15\% ($p < 0.05$) when compared against a conventional shuffled learning strategy on the same motion data. Within each motion category, it either improved or maintained the dice score. To the best of our knowledge, we are the first to quantitatively assess the segmentation bias on various levels of motion artifacts present in a brain MRI image. 

\end{abstract}

\section{Introduction}
\label{intro}

Gliomas are a family of neoplasms of the brain that includes the devastating and most common tumor of the brain, glioblastoma \cite{Marsh2013,Wen2010,Mazzara2004,Bakas2018}. The most aggressive types of gliomas associated with low patient survival rates are the high grade gliomas (HGG), such as glioblastoma, which are extremely infiltrative, spreading extensively through surrounding tissue. Surgery usually focuses on areas of highest grade and debulking of tumor to improve short term symptoms and quality of life \cite{Marsh2013,Bakas2018}. Magnetic Resonance Imaging (MRI) is the standard approach to diagnose and follow gliomas \cite{Wen2010} particularly over the course of their management. Unfortunately, imaging of gliomas, and HGG in particular, is limited by their varied and similar appearance to other neoplasms, irregular shapes, and heterogeneous histologic grade throughout their volume making it difficult to characterize and quantify \cite{Bakas2018,Myronenko2018,Isensee2018,Zhou2019,Zhou2018,Mckinley2018,Muller2019}. Different contrasts, such as T1, T1 with contrast (T1ce), T2, and Fluid Attenuation Inversion Recovery (FLAIR), are usually acquired to visualize the tissue properties and the extent of the tumor \cite{Bakas2018}. Furthermore, varying MRI scanners and exam protocols are used at different institutions to acquire these scans \cite{Bakas2018,Isensee2018}. Given the morbidity surrounding HGGs, there is increasing interest in quantitatively describing these lesions by identifying and segmenting the tumor and its sub-components in MRI images. 

Further confounding analysis is patient motion, the predominant source of image degradation in clinical practice affecting 10\%-42\% of brain exams \cite{Andre2015,Sommer2020}. Degradation of the MR images can potentially be identified at the time of exam, but this would require almost 20\% of all MR exams to be repeated \cite{Andre2015}. In some cases, the entire MR exam can be rejected and the patient will need to be recalled and re-scanned, resulting in tremendous financial and time costs to the healthcare provider \cite{Andre2015}. Moreover, radiologists usually attempt to ``see-through" \cite{Sommer2020} motion artifacts to diagnose the underlying condition, but if the quality of the MR image is severely corrupted, the accurate localization of the tumor boundary is hindered. Many techniques have been proposed in a plethora of prior work to segment brain MR images \cite{Bakas2018}, but convolutional neural network approaches are the state of the art in this domain \cite{Bakas2018,Myronenko2018,Isensee2018,Zhou2019,Zhou2018,Mckinley2018,Muller2019,Hu2020}. Similarly, many schemes have also been proposed for prospective or retrospective \cite{Ooi2009,Pipe1999,Kober2011,Godenscheger2016} motion correction. Deep learning-based approaches for motion correction \cite{Sommer2020,Sommer2018,Duffy2018,Pawar2018,Khalili2019} have also shown promise in improving the image quality. Some approaches have enabled the segmentation performance of a network to improve post-correction \cite{Khalili2019}. We focus on the deep learning approaches in this paper.

The Multimodal Brain Tumor Segmentation Challenge \cite{Bakas2018} has been predominantly used by prior deep learning algorithms, and it provides a dataset of multi-sequence, pre- and post-operative, HGG and low grade gliomas (LGG), neuroradiologist-segmented brain MRIs. These prior lesion segmentation methods utilize all available MR sequences (e.g. T1, T2, T1ce, FLAIR), but in clinical practice, radiologists focus on the slices through the mass of a few sequences (e.g. T1ce or FLAIR) for sizing of the tumor and refer to other sequences for confirmation. Prior research has shown that the addition of a small amount of Gaussian noise to the training data degraded the segmentation performance of the top performing models from the BraTS challenge by 10\%-15\% \cite{Muller2019}. Attempts to make robust motion insensitive models have augmented the training set with images that incorporate simulated motion \cite{Zhou2019,Khalili2019}. But, they do not quantify the algorithmic performance across different severities of motion corruption: minimal, mild, moderate, severe. Thus, the impact and segmentation performance relative to the spectrum of motion artifacts is not characterized. 

In this paper, we assess the bias caused by a gamut of simulated motion on the segmentation performance (in terms of dice score) of neural networks using the BraTS 2019 dataset. We use five previously published neural networks - UNet with Attention \cite{Oktay2018}, UNet 2.5D \cite{Khalili2019}, LSTM UNet \cite{Arbelle2019}, LSTM ResNet34 \cite{Milletari2018}, and LSTM CorNet \cite{Mathai2019} and quantify their segmentation bias. We focus on the task of HGG segmentation in only T1ce sequences affected by different levels of motion corruption. We postulate that a curriculum learning \cite{Bengio2009,Zhou2019} strategy, wherein an easier task (minimal) is first learned followed by progressively harder tasks (mild, moderate, severe), helps these networks converge faster and improves their segmentation performance. The intent is to evaluate and quantify segmentation performance of these algorithms relative to the severity of motion artifact, and show that an alternative learning strategy improves performance.

\noindent
\textbf{Contribution.} 1) For a lesion segmentation task, we characterize the bias caused by different levels of motion artifacts on the segmentation performance of neural networks. 2) We propose a curriculum learning strategy as a way to overcome this bias and improve the segmentation performance. 


\section{Methods}
\label{methods}

\noindent
\textbf{Data.} In this study, the 2019 Brain Tumor Segmentation challenge \cite{Bakas2018} dataset was used for experimentation. The dataset is split into training, validation, and testing sets. The training set contains 259 HGG and 76 LGG subjects whose corresponding ground truth labels were provided. The ground truth masks include four labels: 1 - necrotic (NCR) and non-enhancing (NET) tumor, 2 - edema (ED), 4 - enhancing tumor (ET), and 0 - normal tissue and background. Each case contained four MRI sequences: T1-weighted, T1-weighted contrast-enhanced (T1ce), T2-weighted and FLAIR. The cases were pre-processed before segmentation, including co-registration to an anatomical template, interpolation to an isotropic resolution (1mm$^3$) and skull stripping. The resulting sequences had dimensions of 240 $\times$ 240 $\times$ 155 pixels. In this study, we focus on the tumor core (labels 1 and 4) of the T1ce sequence, which is crucial to clinical tumor size estimation.

\begin{figure}[!h]
\centering
\begin{subfigure}[b]{\columnwidth}
\vspace*{\fill}
  \centering
  \includegraphics[width=\columnwidth,height=4.3cm]{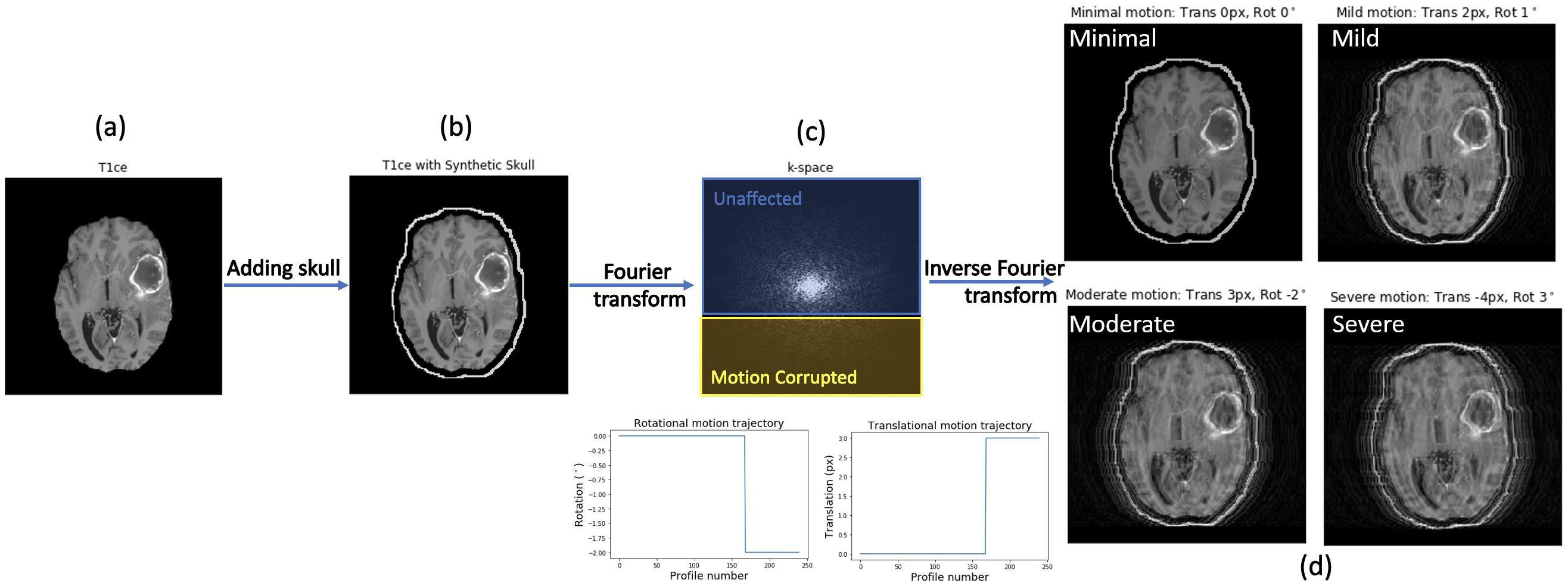}
\end{subfigure} 
\caption{Simulation of rigid body translation and rotation on T1ce images are shown. (a) Original T1ce, (b) T1ce with synthetic skull, (c) Combined k-space signal of unaffected k-space (blue) and motion corrupted k-space (yellow). Example motion trajectories are plotted as a function of k-space profile number. (d) T1ce with four levels of simulated motion artifacts: minimal, mild, moderate and severe. The motion simulation was applied only to the T1ce images, and the underlying ground truth labels were left unchanged.}
\label{fig:motionSimPipeline}
\end{figure}

\noindent 
\textbf{Motion Simulation.} The BraTS dataset is not representative of a real clinical scenario where data is often limited, noisy, and labels are sparse or incorrect \cite{Muller2019}. To mimic a real clinical setting, we utilized only the 259 HGG T1ce sequences from subjects in the training set, and divided them into four motion categories: minimal (64), mild (64), moderate (64), and severe (67). In each motion category, we further subdivided the data into train/validation/testing splits. For the first three motion categories, each split contained 38/6/20 cases, while the split for the severe motion category contained 40/6/21 cases. To study the bias introduced by motion, artifacts were simulated on T1ce sequences assuming a rigid body translation and rotation model as shown in Fig. \ref{fig:motionSimPipeline}. The severity of the artifact depended heavily on the timing and amplitude of the movement during MRI signal sampling. Motion was artificially introduced in the frequency domain (k-space): 1) Minimal (no changes to 64 cases), 2) Mild ($\pm$ 2px translation/$\pm$ 1$^{\circ}$ rotation applied on 64 cases), 3) Moderate ($\pm$ 3px translation/$\pm$ 2$^{\circ}$ rotation applied on 64 cases), 4) Severe ($\pm$ 4px translation/$\pm$ 3$^{\circ}$ rotation applied on 67 cases). The motion corrupted k-space data can be considered a linear combination of two versions of k-space: (1) k-space before motion and (2) k-space after simulated motion event (rotation, translation). When the combined k-space data was converted back as images, varying degrees of simulated motion artifact were evident. To more closely approximate the visual appearance of motion artifact seen in the clinical setting, a simulated skull was added back to the skull-stripped data which produced the typical ringing, blurring and ghosting appearances.


\begin{figure}[!hb]
\centering
\begin{subfigure}[b]{0.24\columnwidth}
\vspace*{\fill}
  \centering
  \includegraphics[width=\columnwidth,height=2.6cm]{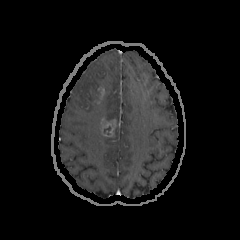}
  \includegraphics[width=\columnwidth,height=2.6cm]{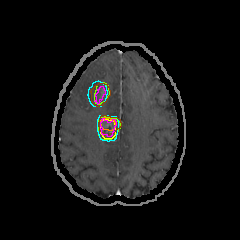}
  \includegraphics[width=\columnwidth,height=2.6cm]{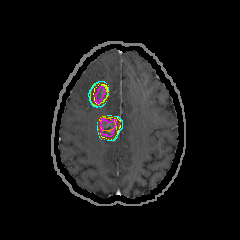}
  \centerline{(a) Minimal}
\end{subfigure} 
\begin{subfigure}[b]{0.24\columnwidth}
\vspace*{\fill}
  \centering
  \includegraphics[width=\columnwidth,height=2.6cm]{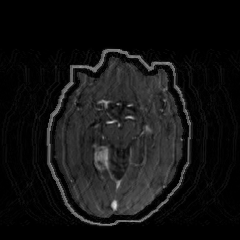}
  \includegraphics[width=\columnwidth,height=2.6cm]{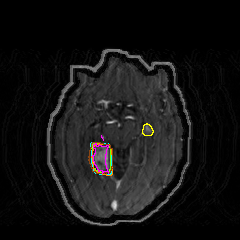}
  \includegraphics[width=\columnwidth,height=2.6cm]{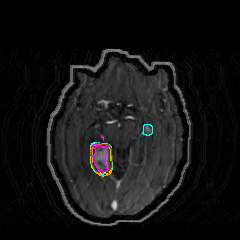}
  \centerline{(b) Mild}
\end{subfigure} 
\begin{subfigure}[b]{0.24\columnwidth}
\vspace*{\fill}
  \centering
  \includegraphics[width=\columnwidth,height=2.6cm]{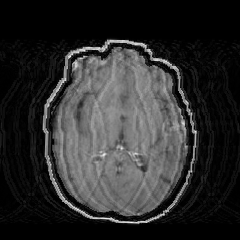}
  \includegraphics[width=\columnwidth,height=2.6cm]{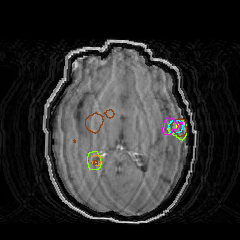}
  \includegraphics[width=\columnwidth,height=2.6cm]{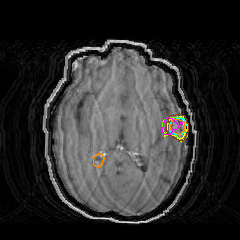}
  \centerline{(c) Moderate}
\end{subfigure} 
\begin{subfigure}[b]{0.24\columnwidth}
\vspace*{\fill}
  \centering
  \includegraphics[width=\columnwidth,height=2.6cm]{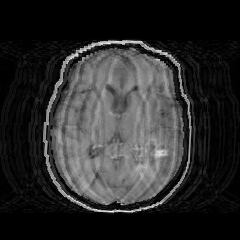}
  \includegraphics[width=\columnwidth,height=2.6cm]{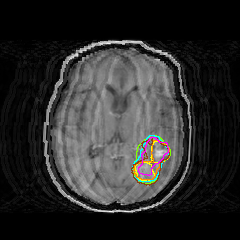}
  \includegraphics[width=\columnwidth,height=2.6cm]{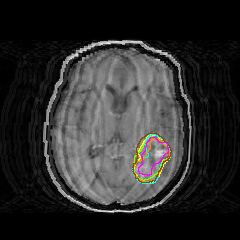}
  \centerline{(d) Severe}
\end{subfigure} 
\caption{Qualitative results of the five different networks on images from test datasets in the (a) minimal, (b) mild, (c) moderate and (d) severe motion categories respectively. The second and third rows show the segmentation results of the networks trained with the shuffled learning and curriculum learning strategies respectively. Magenta: ground truth label, Yellow: UNet Attention, Green: UNet 2.5D, Cyan: LSTM UNet, Orange: LSTM ResNet, Brown: LSTM CorNet.}
\label{fig:qual_images}
\end{figure}

\noindent
\textbf{Baseline Comparisons.} Currently, radiologists center their tumor sizing efforts over a few slices in particular sequences (e.g. T1ce or FLAIR) and refer to the other sequences for confirmation. As prior work has been focused on volumetric brain lesion segmentation, we paid attention to other popular models that utilized either 1- or 3- slice(s) for segmenting the lesion. We extend the prior work by testing the performance of five previously proposed neural networks: 1) UNet with attention \cite{Oktay2018}, 2) UNet 2.5D \cite{Khalili2019}, 3) LSTM UNet \cite{Arbelle2019}, 4) LSTM ResNet34 \cite{Milletari2018}, and 5) LSTM CorNet \cite{Mathai2019}. These network architectures follow the basic encoder-decoder design \cite{Ronneberger2015}. UNet with attention \cite{Oktay2018} used attention gates to identify salient image regions and suppress noisy responses. The LSTM-based networks \cite{Arbelle2019,Milletari2018,Mathai2019} incorporated a convLSTM decoder structure and other modifications, such as residual \cite{He2016} and dilated convolutions \cite{Koltun2016} etc., to exploit the extracted contextual features. The UNet 2.5D \cite{Khalili2019} used three T1ce slices as the network input, while the other networks used a single T1ce slice. 




\noindent
\textbf{Learning Strategy.} Prior work on MR image segmentation has utilized a shuffled learning strategy during training that relied on randomly presenting an individual case to the network from the training set \cite{Bakas2018,Myronenko2018,Isensee2018,Mckinley2018,Muller2019}. Within the context of motion corruption, this strategy does not account for the varying degrees of motion severity. Instead of a randomly shuffled dataset order, a curriculum-based strategy \cite{Bengio2009,Oksuz2019} was used in this work to present the network with easier datasets (e.g. minimal) first, and gradually feeding it with data of increasing motion severity (mild, moderate, severe). Only the order of datasets presented to the network was changed, and the time complexity of learning remained the same as with shuffled learning. 

\section{Experiments}
\label{experiments}

\noindent
\textbf{Design.} We quantitatively evaluated the lesion segmentation performance of each network using the BraTS 2019 data. The first experiment computed the upper limit of performance by using the networks to segment sequences without artifacts or simulated skull (blue bar in Fig. \ref{fig:spread_ls_dc}). Second, the networks were trained and tested on data where the skull was simulated (gold bar in Fig. \ref{fig:spread_ls_dc}). Third, the networks that were trained on artifact-free data containing the simulated skull were tested on artifact simulated data with the skull (green bar in Fig. \ref{fig:spread_ls_dc}). Fourth, the networks were trained on data containing the simulated skull and motion artifacts, and tested on data containing the skull and motion artifacts (orange bar in Fig. \ref{fig:spread_ls_dc}). In the aforementioned experiments, a shuffled learning strategy was used, i.e., all the training data fed to the network were shuffled with no regard to the motion severity. In the last experiment, a curriculum learning strategy was introduced during training time; datasets were fed to the networks during training in the order of increasing motion severity i.e., minimal, mild, moderate, severe. The training data contained both the simulated skull and motion artifacts, and the networks were tested on data containing the skull and motion artifacts (pink bar in Fig. \ref{fig:spread_ls_dc}). 

\noindent
\textbf{Pre-processing.} Each image in the training sequence was normalized \cite{Isensee2018,Zhou2019,Muller2019} by subtracting its mean value and dividing by its standard deviation of the intensities within the brain region. After normalization, the images were padded to 256$\times$256 pixels for training the network. 

\noindent
\textbf{Data Augmentation.} In addition to the motion simulation, heavy data augmentation was conducted in the form of flipping (horizontal and vertical), gamma adjustment, Gaussian noise addition, Gaussian blurring, Median blurring, Bilateral blurring, cropping, and affine transformations.

\noindent
\textbf{Training.} A grid search across the parameter space was conducted for each network on a 4$\times$ sub-sampled dataset to identify the best hyper-parameters given the input data. We trained each network for 30 epochs with the training being terminated early if the validation loss did not improve for 7 epochs. The networks were trained with the dice loss \cite{Milletari2018} as it has been experimentally proven to be less sensitive to the class imbalance problem. In most cases, grid search yielded chosen parameters that were the same as those proposed in the original papers with the exception of the batch size, which varied from 4-16 T1ce images per batch. The reader is referred to the original papers for full implementation details. The ADAM optimizer \cite{Kingma2015} was used for optimizing the loss function, and the final pixel level probabilities were classified using the softmax function. All experiments were run on a workstation running Ubuntu 16.04LTS, a NVIDIA Titan Xp GPU, and the average inference time on a test slice was 44.7ms for UNet Attention, 49.1ms for UNet 2.5D, 58.8ms for LSTM UNet, 86.3ms for LSTM ResNet34, and 96.2ms for LSTM CorNet.

\noindent
\textbf{Post-processing.} In contrast to the prior approaches, no post-processing was done on the predicted segmentation masks. The predicted labels were directly compared against the ground truth masks, and the dice score was computed. 

\section{Results and Discussion}
\label{expRes}

\begin{figure}[!ht]
\centering
\begin{tabular}{cccc}
\includegraphics[width=.45\textwidth,height=4.5cm]{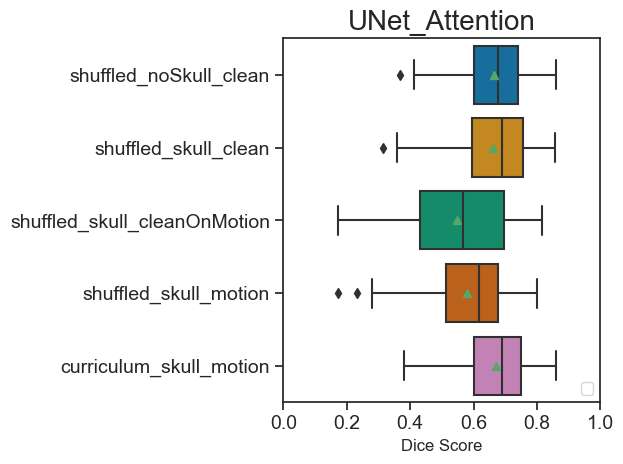} &
\includegraphics[width=.45\textwidth,height=4.5cm]{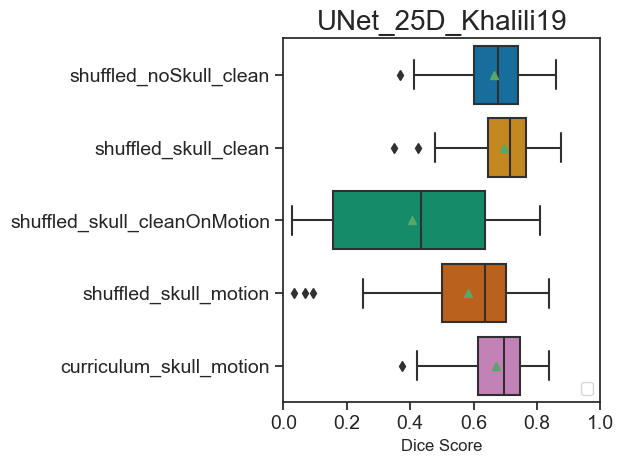} \\
\textbf{(a)}  & \textbf{(b)}  \\
\end{tabular}
\begin{tabular}{cccc}
\includegraphics[width=.3\textwidth,height=3.5cm]{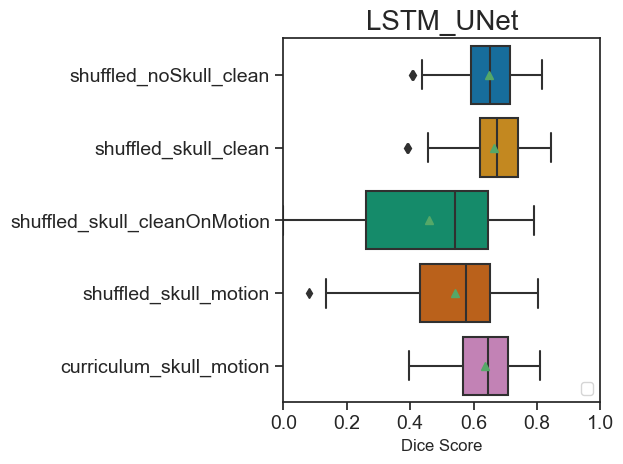} &
\includegraphics[width=.3\textwidth,height=3.5cm]{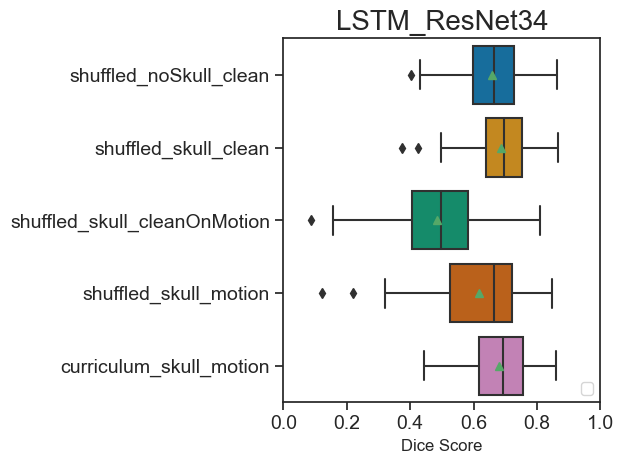} &
\includegraphics[width=.3\textwidth,height=3.5cm]{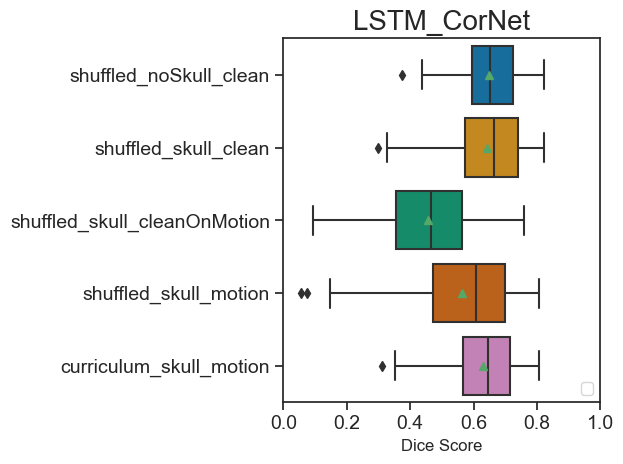} \\
\textbf{(c)}  & \textbf{(d)} & \textbf{(e)}  \\
\end{tabular}
\caption{Each plot displays the spread of the dice scores obtained using different learning strategies and data collections across all the testing set for the (a) UNet with attention model, (b) UNet 2.5D model, (c) LSTM UNet model, (d) LSTM ResNet34 model, (e) LSTM CorNet model.}
\label{fig:spread_ls_dc}
\end{figure}

\noindent
\textbf{Results across all motion categories.} As seen in Fig. \ref{fig:qual_images}, qualitative results are presented for all the models on test dataset images organized by motion category. Shuffled learning tends to under-segment the tumor (false negatives) and in some cases generates false positives, contributing to a lower dice score. The spread of the dice scores across the test datasets for the five networks are presented in Fig. \ref{fig:spread_ls_dc}, and the associated mean and standard deviations of the dice scores are listed in Table \ref{diceErrorComparison_dice_LS_DC}. From the experimental designs in Sec. \ref{experiments}, these networks have been trained with different learning strategies and data categories. Table \ref{pErrorComparison_p_LS} lists the $p$-values from statistical analysis using ANOVA that was conducted on dice scores between different learning strategies and motion categories for each network.

From Fig. \ref{fig:spread_ls_dc}, adding the simulated skull to the data did not change the segmentation performance when compared against data without the skull across the neural networks (blue vs gold bars, p$>$0.05). However, the performance degraded significantly when a network trained with the shuffled learning strategy and clean data was tested on motion data (gold vs green bars, p$<${}$<$0.05); e.g., the dice scores of the UNet Attention network dropped by $>$12\% when the network trained on clean data was tested on motion corrupted data. This observation holds true across the four other networks, as shown in Table \ref{diceErrorComparison_dice_LS_DC} and Fig. \ref{fig:spread_ls_dc}. Also consistent with prior work \cite{Khalili2019}, a 23\% mean improvement in performance across all networks was seen with incorporating motion data during training. Fig. \ref{fig:spread_ls_dc} also demonstrates that curriculum learning compensates for the motion corruption. From Table \ref{diceErrorComparison_dice_LS_DC}, the results indicate that the curriculum learning strategy provided a 9\%-15\% improvement when compared against the shuffled learning strategy on the same motion data (orange vs pink bars, $p < 0.05$). Across all five neural networks, the resulting segmentations from curriculum learning were comparable to the corresponding cases where the networks were trained on clean data (gold vs pink bars, $p > 0.05$) as seen from the dice scores in Table \ref{diceErrorComparison_dice_LS_DC} and $p$-values in Table \ref{pErrorComparison_p_LS} (last row).

\begin{figure}[!ht]
\centering
\begin{subfigure}[b]{0.192\columnwidth}
\vspace*{\fill}
  \centering
  \includegraphics[width=\columnwidth,height=2.5cm]{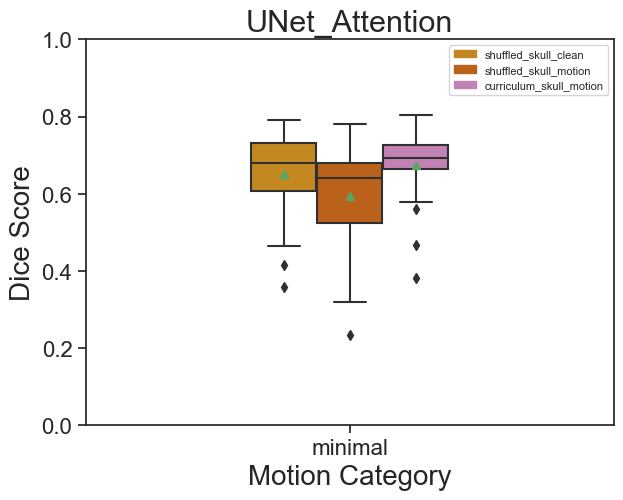}
  \centerline{}
  \includegraphics[width=\columnwidth,height=2.5cm]{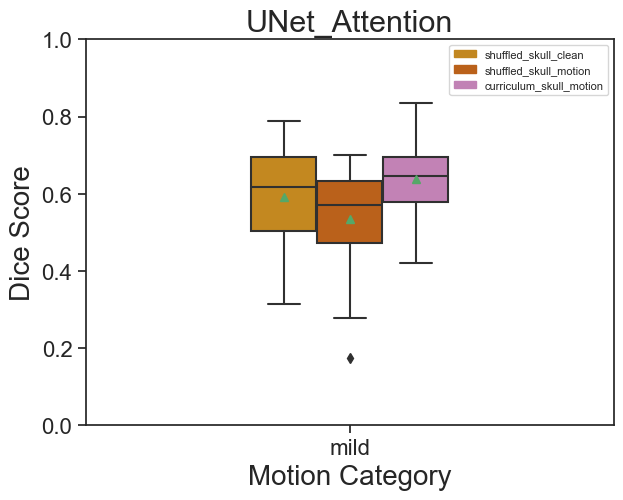}
  \centerline{}
  \includegraphics[width=\columnwidth,height=2.5cm]{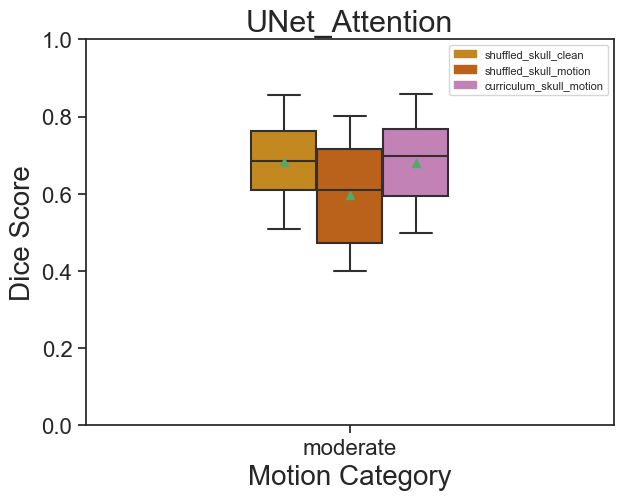}
  \centerline{}
  \includegraphics[width=\columnwidth,height=2.5cm]{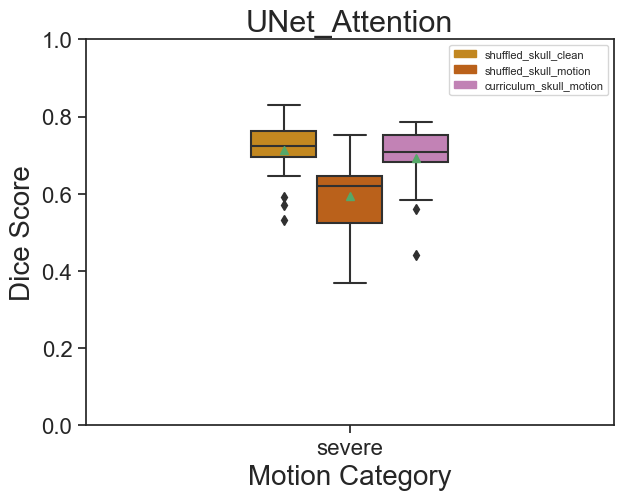}
  \centerline{(a)}
\end{subfigure} 
\begin{subfigure}[b]{0.192\columnwidth}
\vspace*{\fill}
  \centering
  \includegraphics[width=\columnwidth,height=2.5cm]{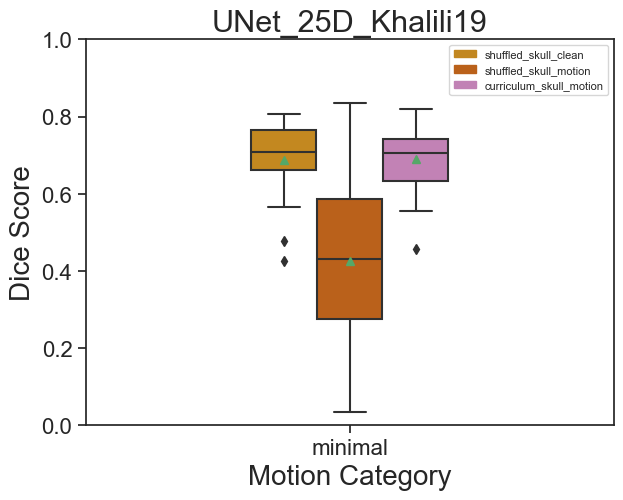}
  \centerline{}
  \includegraphics[width=\columnwidth,height=2.5cm]{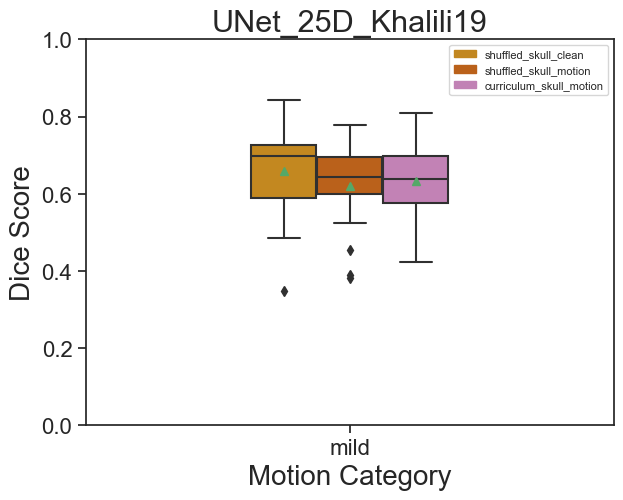}
  \centerline{}
  \includegraphics[width=\columnwidth,height=2.5cm]{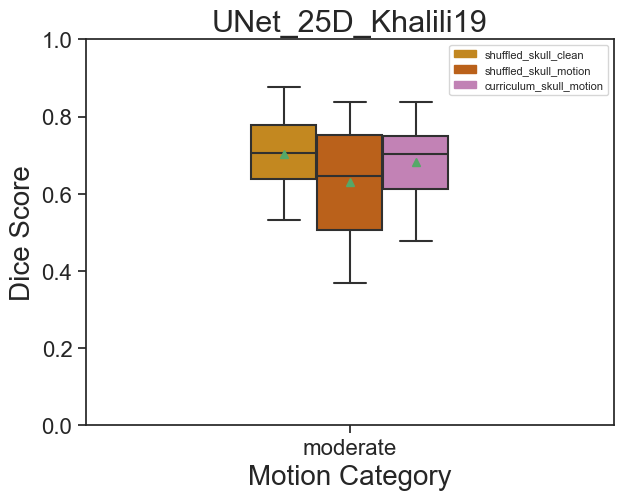}
  \centerline{}
  \includegraphics[width=\columnwidth,height=2.5cm]{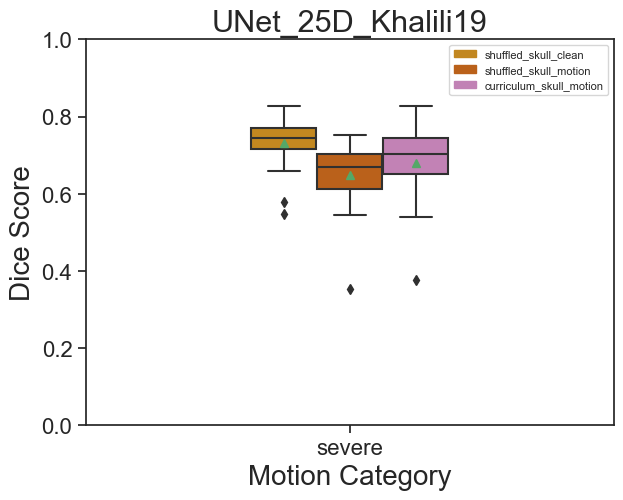}
  \centerline{(b)} 
\end{subfigure} 
\begin{subfigure}[b]{0.192\columnwidth}
\vspace*{\fill}
  \centering
  \includegraphics[width=\columnwidth,height=2.5cm]{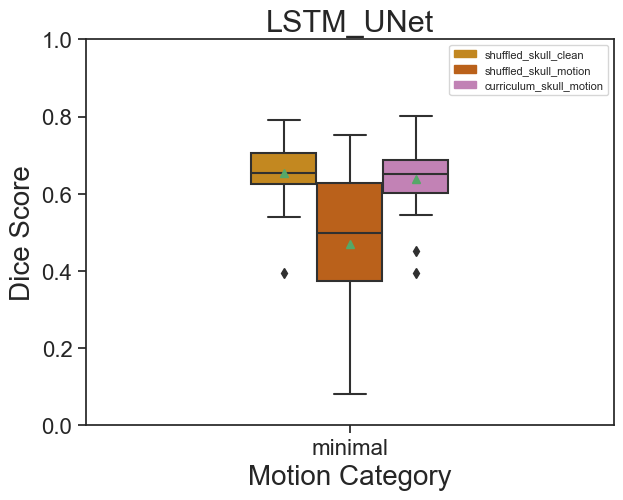}
  \centerline{}
  \includegraphics[width=\columnwidth,height=2.5cm]{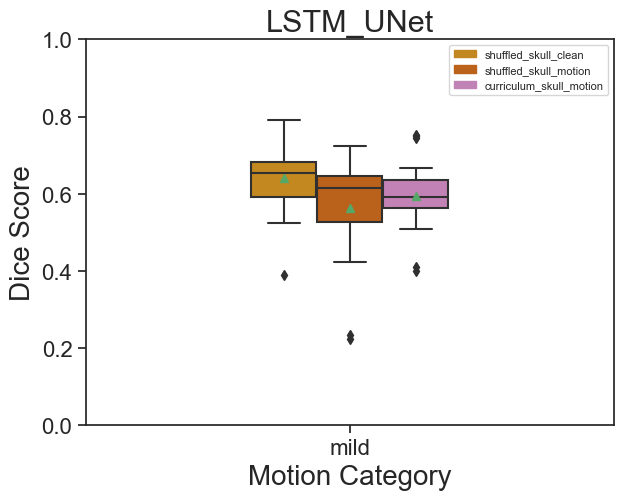}
  \centerline{}
  \includegraphics[width=\columnwidth,height=2.5cm]{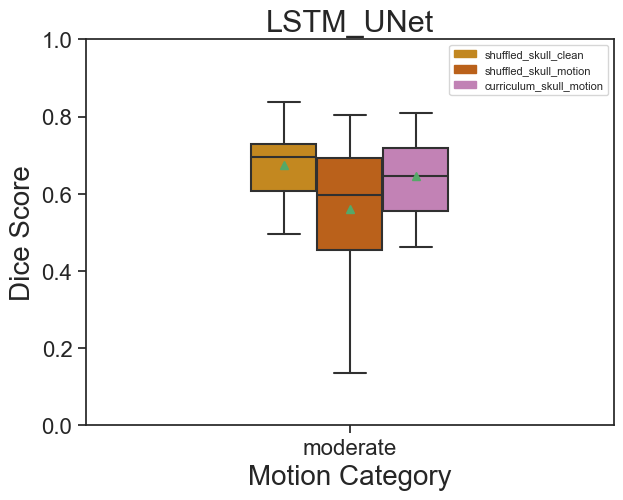}
  \centerline{}
  \includegraphics[width=\columnwidth,height=2.5cm]{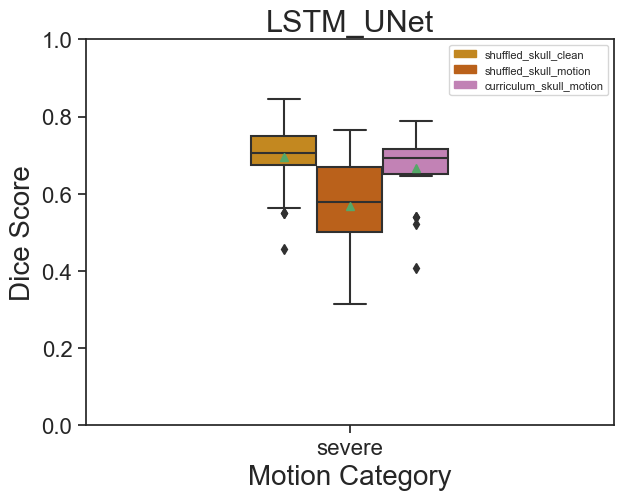}
  \centerline{(c)} 
\end{subfigure} 
\begin{subfigure}[b]{0.192\columnwidth}
\vspace*{\fill}
  \centering
  \includegraphics[width=\columnwidth,height=2.5cm]{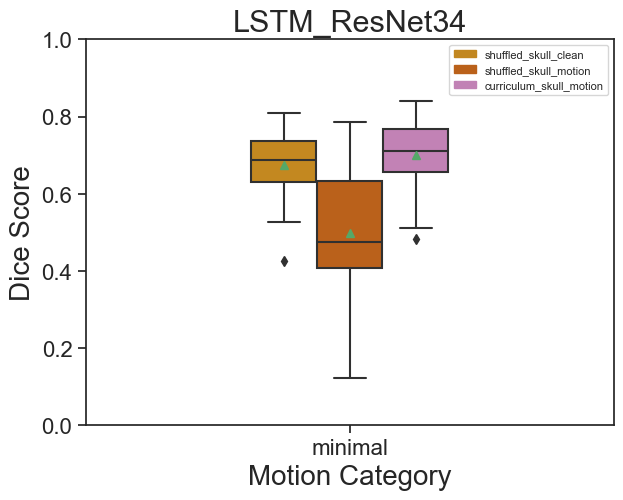}
  \centerline{}
  \includegraphics[width=\columnwidth,height=2.5cm]{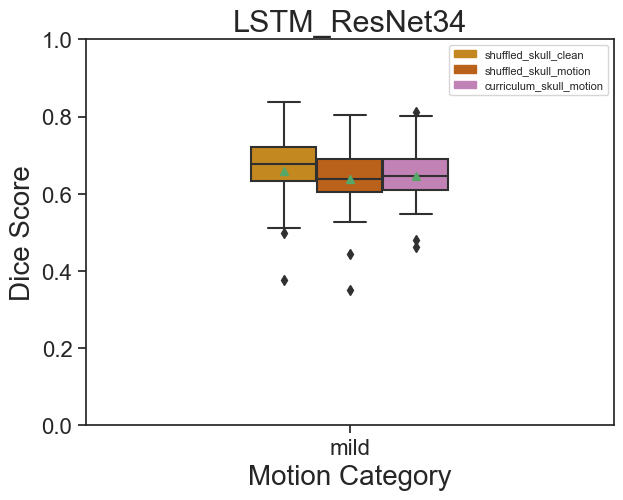}
  \centerline{}
  \includegraphics[width=\columnwidth,height=2.5cm]{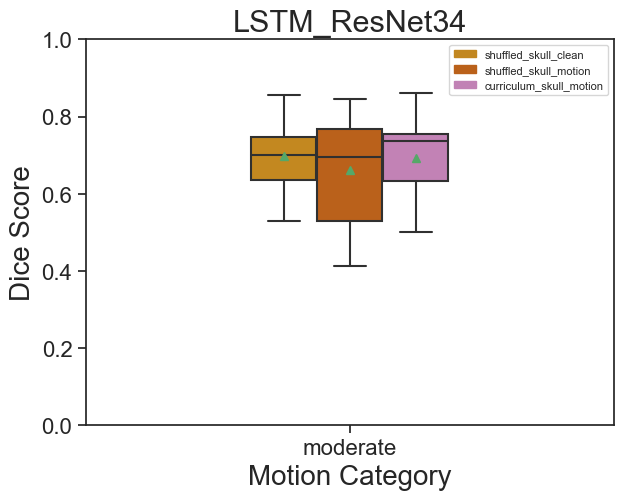}
  \centerline{}
  \includegraphics[width=\columnwidth,height=2.5cm]{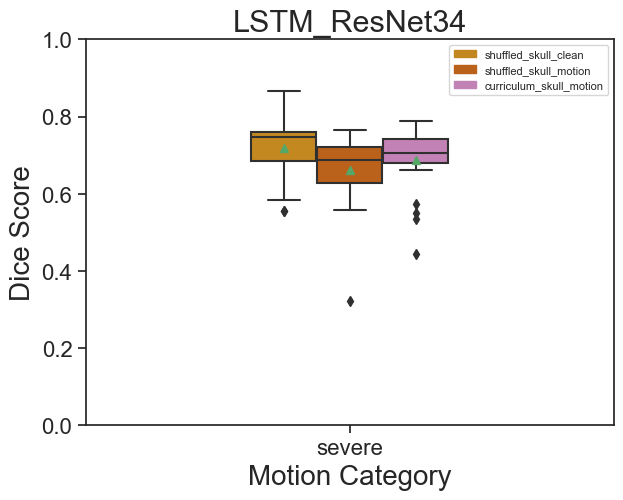}
  \centerline{(d)} 
\end{subfigure} 
\begin{subfigure}[b]{0.192\columnwidth}
\vspace*{\fill}
  \centering
  \includegraphics[width=\columnwidth,height=2.5cm]{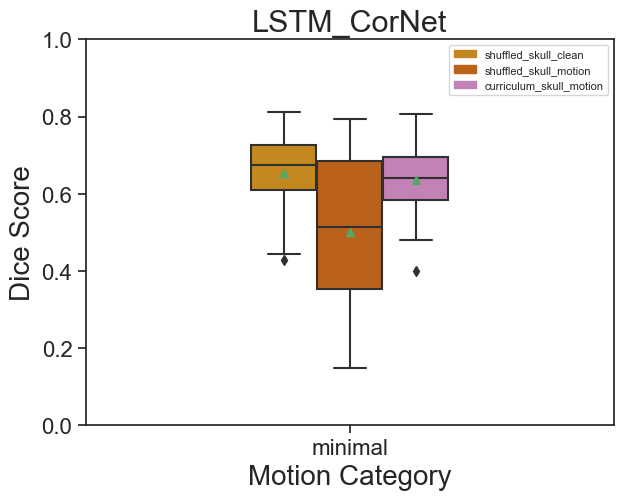}
  \centerline{}
  \includegraphics[width=\columnwidth,height=2.5cm]{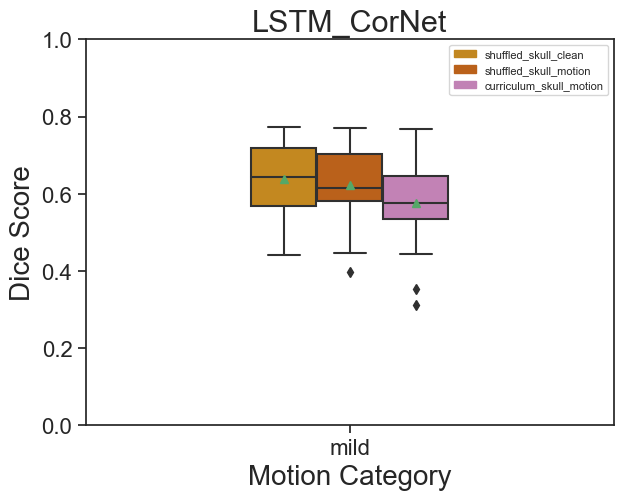}
  \centerline{}
  \includegraphics[width=\columnwidth,height=2.5cm]{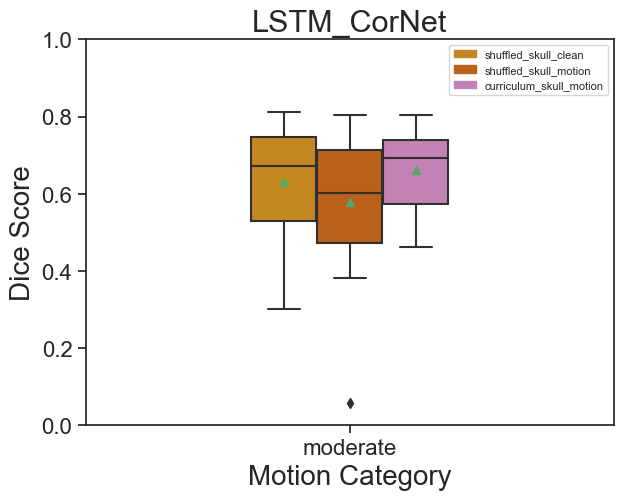}
  \centerline{}
  \includegraphics[width=\columnwidth,height=2.5cm]{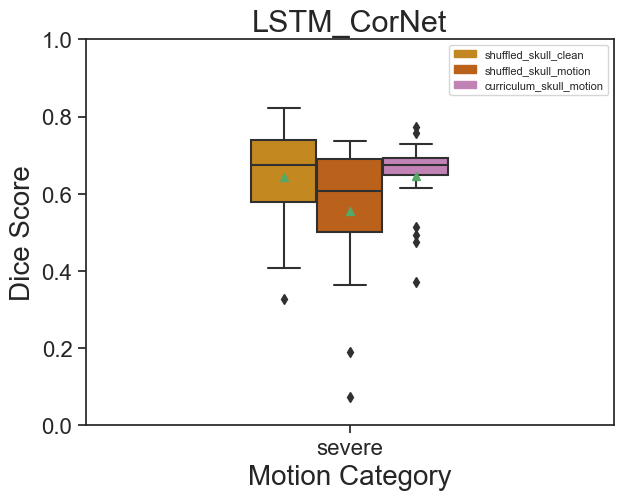}
  \centerline{(e)} 
\end{subfigure}
\caption{Dice scores for the (a) UNet with attention, (b) UNet 2.5D, (c) LSTM UNet, (d) LSTM ResNet34, (e) LSTM CorNet across the various motion categories. The first two error bars show the shuffled learning strategy trained with/without motion corrupted data respectively, and the last error bar displays the curriculum learning strategy trained with motion corrupted data. Curriculum learning either enhanced or maintained the segmentation performance across the different motion categories.}
\label{fig:spread_ls_moco_comparison}
\end{figure}

\begin{figure}[!h]
\centering
\begin{subfigure}[b]{0.25\columnwidth}
\vspace*{\fill}
  \centering
  \includegraphics[width=\columnwidth,height=2.5cm]{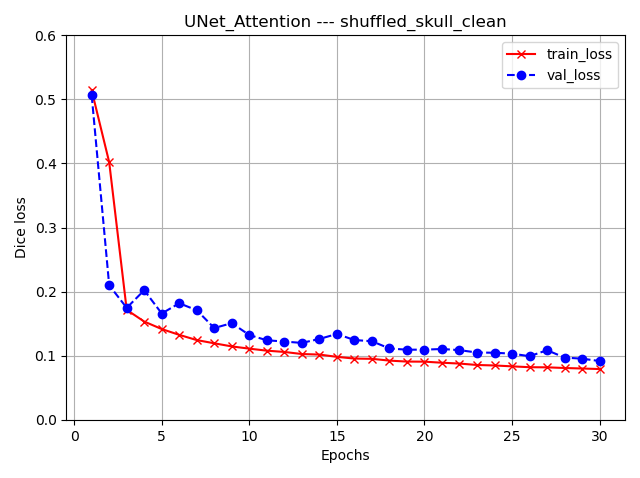}
  \centerline{(a)}
\end{subfigure} 
\begin{subfigure}[b]{0.25\columnwidth}
\vspace*{\fill}
  \centering
  \includegraphics[width=\columnwidth,height=2.5cm]{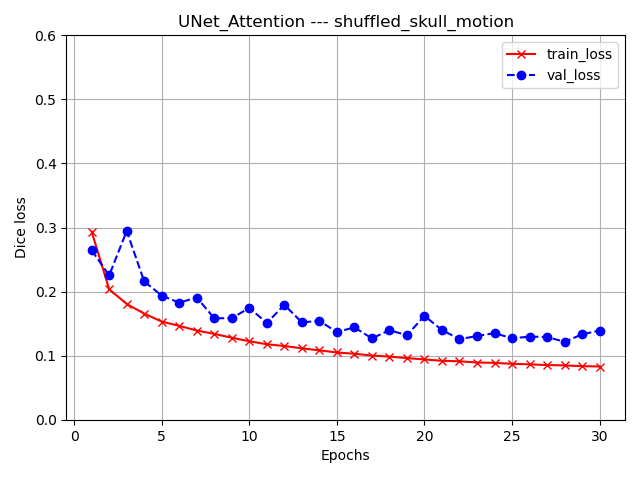}
  \centerline{(b)} 
\end{subfigure} 
\begin{subfigure}[b]{0.25\columnwidth}
\vspace*{\fill}
  \centering
  \includegraphics[width=\columnwidth,height=2.5cm]{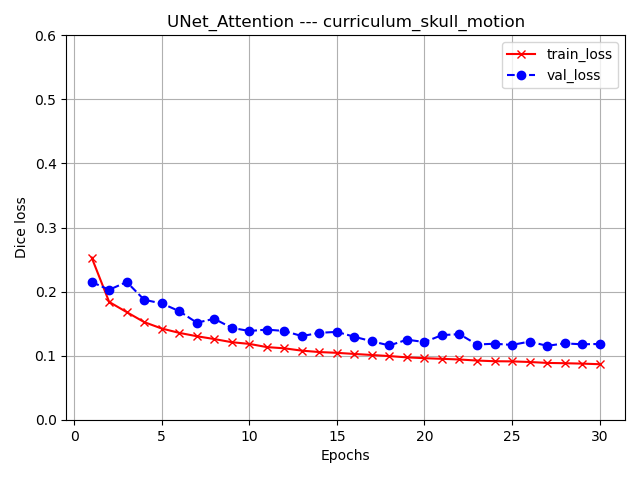}
  \centerline{(c)} 
\end{subfigure} 
\caption{Comparison of the training loss (red) and validation loss (blue) for the UNet architecture with attention mechanism with (a) shuffled learning strategy on clean data, (b) shuffled learning strategy on motion data, and (c) curriculum learning strategy on motion data. }
\label{fig:trainValLoss}
\end{figure}

\noindent
\textbf{Results for each motion category.} Fig. \ref{fig:spread_ls_moco_comparison} shows the performance of networks trained with different learning strategies on the individual motion categories. For each network, the dice scores of networks trained with a shuffled learning strategy on clean data are also shown for reference. Table \ref{pErrorComparison_p_moCats} shows the statistical significance of the two learning strategies on each motion category; following a test of normality, either a paired t-test or the Wilcoxon signed rank test was used. Fig. \ref{fig:trainValLoss} depicts the training loss and the validation loss as a function of epochs for the UNet Attention network: (a) shuffled learning strategy on clean data, (b) shuffled learning strategy on motion data, and (c) curriculum learning strategy on motion data. In contrast to the shuffled learning, curriculum learning exhibits a relatively stable convergence rate with a lower loss. Other observations that were made for each motion category include:

\begin{itemize}

\item \textbf{UNet Attention}: From Fig. \ref{fig:spread_ls_moco_comparison}(a) and Table \ref{pErrorComparison_p_moCats}, curriculum learning improved the segmentation performance across motion categories (p $<$ 0.05). 
\item \textbf{UNet 2.5D}: From Fig. \ref{fig:spread_ls_moco_comparison}(b) and Table \ref{pErrorComparison_p_moCats}, curriculum learning improved the dice score significantly for the minimal, moderate and severe category (p $<$ 0.05). No significant difference was found between the two learning strategies for the mild motion category (p = 0.257). Curriculum learning either improved or maintained the segmentation performance by yielding higher or comparable dice scores.
\item \textbf{LSTM UNet}: As seen in Table \ref{pErrorComparison_p_moCats}, curriculum learning improved the segmentation performance across all motion categories, except for mild motion. However, it resulted in lower standard deviations in the mild category in contrast to shuffled learning as shown in Fig. \ref{fig:spread_ls_moco_comparison}(c). 
\item \textbf{LSTM ResNet34}: As seen in Fig. \ref{fig:spread_ls_moco_comparison}(d) and Table \ref{pErrorComparison_p_moCats}, significantly higher dice scores were obtained with curriculum learning for minimal and severe motion categories (p $<$ 0.05), but there is no statistically significant difference with the other motion groups. Curriculum learning did provide comparable dice scores in the mild motion category, and a higher mean dice score and a lower standard deviation in the moderate category. 
\item \textbf{LSTM CorNet}: Curriculum learning improved the segmentation performance across all motion categories as shown in Fig. \ref{fig:spread_ls_moco_comparison}(e) and Table \ref{pErrorComparison_p_moCats}. 
\end{itemize}

\begin{table}[t]
\centering\fontsize{9}{12}\selectfont 
\setlength\aboverulesep{0pt}\setlength\belowrulesep{0pt} 
\setlength{\tabcolsep}{7pt} 
\setcellgapes{3pt}\makegapedcells 
\caption{Comparison of the dice score for each neural network trained with different learning strategies and data collections.}
\begin{adjustbox}{max width=\textwidth}
\begin{tabular}{@{} c|c|c|c|c|c @{}} 
\toprule
                        & UNet Attention     & UNet 2.5D         & LSTM UNet         & LSTM ResNet34       & LSTM CorNet       \\
\midrule
Shuffled noSkull Clean          & 0.67 $\pm$ 0.1    & 0.67 $\pm$ 0.1    & 0.65 $\pm$ 0.09   & 0.66 $\pm$ 0.1    & 0.65 $\pm$ 0.09    \\
Shuffled Skull Clean            & 0.66 $\pm$ 0.12   & 0.7 $\pm$ 0.1     & 0.67 $\pm$ 0.09   & 0.69 $\pm$ 0.09   & 0.64 $\pm$ 0.12    \\
Shuffled Skull CleanOnMotion    & 0.55 $\pm$ 0.16   & 0.4 $\pm$ 0.24    & 0.46 $\pm$ 0.22   & 0.48 $\pm$ 0.17   & 0.46 $\pm$ 0.14    \\ 
Shuffled Skull Motion           & 0.58 $\pm$ 0.13   & 0.58 $\pm$ 0.17   & 0.54 $\pm$ 0.16   & 0.62 $\pm$ 0.15   & 0.56 $\pm$ 0.17    \\ 
Curriculum Skull Motion         & 0.67 $\pm$ 0.1    & 0.67 $\pm$ 0.1    & 0.64 $\pm$ 0.1    & 0.68 $\pm$ 0.09   & 0.63 $\pm$ 0.11    \\ 
\bottomrule
\end{tabular}
\end{adjustbox}
\label{diceErrorComparison_dice_LS_DC}
\end{table}
\begin{table}[!t]
\centering\fontsize{9}{12}\selectfont 
\setlength\aboverulesep{0pt}\setlength\belowrulesep{0pt} 
\setlength{\tabcolsep}{7pt} 
\setcellgapes{3pt}\makegapedcells 
\caption{Statistical significance of networks trained with different learning strategies across all motion categories at a 5\% significance level. \\\hspace{\textwidth} s = shuffled, c = curriculum, S = Skull, C = Clean, M = Motion, n = no} 
\begin{adjustbox}{max width=\textwidth}
\begin{tabular}{@{} c|c|c|c|c|c @{}} 
\toprule
            & UNet Attention     & UNet 2.5D         & LSTM UNet         & LSTM ResNet34       & LSTM CorNet       \\
\midrule
$s_{nSC}$ \quad vs. \quad $s_{SC}$   & 0.665     & 0.068    & 0.211     & 0.058    & 0.724    \\
$s_{SC}$ \quad vs. \quad $s_{SM}$    & 0.0001     & 0.0001     & 0.0007     & 0.001     & 0.001    \\
$s_{SM}$ \quad vs. \quad $c_{SM}$    & 0.0001     & 0.0001     & 0.0001     & 0.001     & 0.004    \\ 
$s_{SC}$ \quad vs. \quad $c_{SM}$    & 0.525     & 0.125     & 0.059     & 0.688     & 0.508    \\ 
\bottomrule
\end{tabular}
\end{adjustbox}
\label{pErrorComparison_p_LS}
\end{table}


\begin{table}[!t]
\centering\fontsize{9}{12}\selectfont 
\setlength\aboverulesep{0pt}\setlength\belowrulesep{0pt} 
\setlength{\tabcolsep}{7pt} 
\setcellgapes{3pt}\makegapedcells 
\caption{Statistical comparison of the learning strategies (shuffled vs. curriculum) used for training different networks on each motion category with $\alpha = 0.05$.}
\begin{adjustbox}{max width=\textwidth}
\begin{tabular}{@{} c|c|c|c|c|c @{}} 
\toprule
            & UNet Attention     & UNet 2.5D         & LSTM UNet         & LSTM ResNet34       & LSTM CorNet       \\
\midrule
Minimal     & 0.0005     & 4.91e-5    & 0.0002     & 71e-5    & 0.0029 \\
Mild        & 8.81e-6     & 0.257     & 0.232     & 0.473     & 0.0012 \\
Moderate    & 8.41e-8     & 0.007     & 0.003     & 0.056     & 0.0045 \\ 
Severe      & 6.81e-5     & 0.012     & 6.81e-5     &  0.038     & 0.0003  \\ 
\bottomrule
\end{tabular}
\end{adjustbox}
\label{pErrorComparison_p_moCats}
\end{table}

\medskip
\noindent
\textbf{Discussion.} From Fig. \ref{fig:spread_ls_dc} and Tables \ref{diceErrorComparison_dice_LS_DC} and \ref{pErrorComparison_p_LS}, when contrasted against curriculum learning, shuffled learning on motion corrupted data resulted in significantly inferior segmentation performance. Potential reasons include an unbalanced set of data for training (64 cases without added motion; 195 cases with added motion), and weight oscillation during training. From Fig. \ref{fig:trainValLoss}, curriculum learning overcame the weight oscillations \cite{Bengio2009} with the data being fed to a network in the order of increasing motion corruption. Overall, the curriculum learning strategy helped to improve the segmentation dice score for all the neural networks that were tested. An in-depth analysis suggested that curriculum learning either improved or maintained the segmentation dice score across the different motion categories for all the networks. While the least improvements in dice scores were seen with the mild motion category, improvements were visualized in the other (minimal, moderate and severe) motion categories across all the networks tested. Our results suggest that the curriculum learning is most beneficial to the UNet with Attention model. 

In general, there is a tendency of curriculum learning to overestimate the volume of the mass as illustrated in Fig. \ref{fig:qual_images}. While a T1ce image was corrupted by simulating motion using rotations, translations, blurring, and ringing artifacts, the underlying ground truth label was not altered in the same way; due to MR physics, applying the same motion simulation to the ground truth label would not ascertain a 1-1 correspondence with the image. Thus, there was a slight mismatch between the motion corrupted image and its corresponding ground truth label. As a network is fed these images during training and minimizes the dice loss function, it has to consider the label mismatch. Therefore, it accounted for the motion and over-segmented the lesion by generating false positives, such that the predicted labels are slightly larger than the underlying ground truth. This caused a drop in the dice scores. Moreover, the dice scores from our study are not as high as those from the BraTS challenge studies, possibly due to training being done on a single image contrast with only 1/3 slice(s). Furthermore, no additional post-processing methods were performed to rectify the predicted segmentation, which has been shown to significantly enhance the dice scores \cite{Bakas2018,Myronenko2018,Isensee2018,Zhou2019,Zhou2018,Mckinley2018,Muller2019}. We would like to point out that the main purpose of this paper was not to achieve the highest segmentation performance. Instead, these results help to quantify the effects of motion corruption, which is commonly encountered in clinical imaging, on the segmentation performance of a variety of common deep learning models.
 
Our observation suggests that for a network trained on motion corrupted data for the task of lesion segmentation and not motion correction, it might be necessary to shrink the prediction while interpreting clinical cases. Potentially, conventional image analysis techniques could be used to erode the labels further down to the lesion margin. While human vision is different and may be tempered by a variety of other information than just the pixels on the screen, our work provides some insight into the effect of motion artifacts on the radiologist interpretations of lesion size. 3D volumetric techniques have been shown to more accurately assess changes in glioblastoma \cite{Dempsey2005}, but it is rarely performed due to the cumbersome segmentation process and limited tools. Furthermore, as radiologists size lesions often over the course of a busy clinical day, there is a need for accurate 2D/2.5D automated segmentation techniques. If automated techniques are to be applied to generate volumetric measurements of lesions, our results show that models will need adequate experience with motion corrupted data. Since motion is an ubiquitous issue in clinical imaging, it will also behoove those analyzing model performance to ensure that results are robust to motion artifacts, or at least identify it as a factor and fail gracefully.

\section{Conclusion and Future Work}

Volumetric assessment of neoplasms are more accurate and clinically relevant. However, in a busy clinical practice, this is rarely performed due to limited tools and the tedious mass segmentation process. Automated tools can fill this need, but they need to be robust to the commonplace artifacts degrading clinical imaging, such as the ubiquitous motion artifact. In this paper, we analyzed the performance of neural networks tasked with segmenting lesions in motion corrupted brain MRI images, and quantified the bias exhibited by them on individual motion categories. We also explored a different learning strategy in curriculum learning, and demonstrated that it is effective in compensating for the effects of simulated motion. It either improved or maintained the segmentation performance across all five networks and four motion severity categories. To the best of our knowledge, this is the first quantitative assessment of segmentation bias from motion artifacts on brain MRI tumor measurement. In the future, we plan to correct for the motion and segment the MR image in a multi-task fashion, and explore curriculum learning further with bootstrapping. 



\section*{Acknowledgements}

We would like to thank Karsten Sommer, Axel Saalbach, Ekta Walia, Chris Martel, Prashanth Pai, and Shawn Stapleton for their comments during discussions related to this work.



\begin{thebibliography}{4}
\small

\bibitem{Marsh2013} J. Marsh et al. ``Current status of immunotherapy and gene therapy for high-grade gliomas", J Clin. Oncology, 20(1), pp. 43-48, (2013).

\bibitem{Wen2010} P. Wen et al. ``Updated response assessment criteria for high-grade gliomas: response assessment in neuro-oncology working group", J Clin. Oncology, 28(11), pp. 1963-1972, (2010).

\bibitem{Mazzara2004} G. Mazzara et al. ``Brain tumor target volume determination for radiation treatment planning through automated MRI segmentation", Int. J Rad. Oncology, Biology, Physics, 59(1), pp. 300-312, (2004).

\bibitem{Bakas2018} S. Bakas et al., ``Identifying the Best Machine Learning Algorithms for Brain Tumor Segmentation, Progression Assessment, and Overall Survival Prediction in the BRATS Challenge", arXiv, (2018).

\bibitem{Myronenko2018} A. Myronenko, ``3D MRI brain tumor segmentation using autoencoder regularization", BrainLes Workshop, MICCAI, (2018).

\bibitem{Isensee2018} F. Isensee et al., ``No New-Net", BrainLes Workshop, MICCAI BrainLes 2018.

\bibitem{Zhou2019} C. Zhou et al., ``Learning Contextual and Attentive
Information for Brain Tumor Segmentation", BrainLes Workshop, MICCAI, pp. 497-507, (2018).

\bibitem{Zhou2018} C. Zhou et al., ``One-pass multi-task convolutional neural networks for efficient brain tumor segmentation", MICCAI, pp. 637-645, (2018).

\bibitem{Mckinley2018} R. McKinley et al., ``Ensembles of Densely-Connected CNNs with Label-Uncertainty for Brain Tumor Segmentation", BrainLes Workshop, MICCAI, (2018).

\bibitem{Muller2019} S. Muller et al., ``Robustness of Brain Tumor Segmentation", arXiv, (2019). 

\bibitem{Hu2020} X. Hu et al., ``Brain SegNet: 3D local refinement network for brain lesion segmentation", BMC Med. Imaging, 20(17), (2020). 

\bibitem{Andre2015} J. Andre et al., ``Toward quantifying
the prevalence, severity, and cost associated with patient
motion during clinical MR examinations", J Am Coll Radiology, 12, pp. 689-695, (2015). 

\bibitem{Ooi2009} M. Ooi et al., ``Prospective real-time correction for arbitrary head motion using active markers", Mag. Res. Med., 62(4), pp. 943-954, (2009).

\bibitem{Pipe1999} J. Pipe et al., ``Motion correction with PROPELLER MRI: application to head motion and free-breathing cardiac imaging", Mag. Res. Med., 42, pp. 963-969, (1999).

\bibitem{Kober2011} T. Kober et al., ``Head motion detection using FID navigators", Mag. Res. Med., 66(1), pp. 135-143, (2011).

\bibitem{Godenscheger2016} F. Godenschweger et al., ``Motion correction in MRI of the Brain", Phys. Med. Biol., 61(5), (2018). 

\bibitem{Sommer2020}, K. Sommer et al., ``Correction of Motion Artifacts Using a Multiscale Fully Convolutional Neural Network", AJNR, (2020). 

\bibitem{Sommer2018} K. Sommer et al., ``Correction of motion artifacts
using a multi-resolution fully convolutional neural network", ISMRM, (2018). 

\bibitem{Duffy2018} B. Duffy et al., ``Retrospective correction of
motion artifact affected structural MRI images using deep learning
of simulated motion", MIDL, (2018). 

\bibitem{Pawar2018} K. Pawar et al., ``Motion correction in MRI using deep convolutional neural network", ISMRM, (2018). 

\bibitem{Khalili2019} N. Khalili et al., ``Generative Adversarial Network for Segmentation of Motion Affected Neonatal Brain MRI", MICCAI, pp. 320-328, (2019). 

\bibitem{Shaw2020} R. Shaw et al., ``A k-Space Model of Movement Artefacts: Application to Segmentation Augmentation and Artefact Removal", IEEE Trans. Med. Imaging, (2020). 

\bibitem{Bengio2009} Y. Bengio et al., ``Curriculum Learning", ICML, pp. 41-48, (2009). 

\bibitem{ChenChen2020} C. Chen et al., ``Realistic Adversarial Data Augmentation for
MR Image Segmentation", arXiv, (2020). 

\bibitem{Ronneberger2015} O. Ronneberger et al., ``U-Net: Convolutional Networks for Biomedical Image Segmentation", MICCAI, 9351, (2015).

\bibitem{Oktay2018} O. Oktay et al., ``Attention U-Net: Learning Where to Look for the Pancreas", MIDL, (2018). 

\bibitem{Arbelle2019} S. Arbelle et al., ``Microscopy Cell Segmentation via Convolutional LSTM Networks", IEEE ISBI, pp. 1008-1012, (2019).

\bibitem{Milletari2018} F. Milletari et al., ``CFCM: Segmentation via Coarse to Fine Context Memory", MICCAI 11073, (2018).

\bibitem{Mathai2019} T. Mathai et al., ``Segmentation of Vessels in Ultra High Frequency Ultrasound Sequences Using Contextual Memory", MICCAI, pp. 173-181, (2019).

\bibitem{He2016} K. He et al., ``Deep Residual Learning for Image Recognition", IEEE CVPR, pp. 770-778, (2016).

\bibitem{Koltun2016} V. Koltun et al., ``Multi-Scale Context Aggregation by Dilated Convolutions", ICLR, (2016).

\bibitem{Oksuz2019} I. Oksuz et al., ``Automatic CNN-based detection of cardiac MR motion artefacts using k-space data augmentation and curriculum learning", Med. Image Anal., 55, pp. 136-147, (2019). 

\bibitem{Kingma2015} D. Kingma et al., ``Adam: a Method for Stochastic Optimization'', ICLR, (2015).

\bibitem{Dempsey2005} M. Dempsey, ``Measurement of Tumor “Size” in Recurrent Malignant Glioma: 1D, 2D, or 3D?", AJNR, 26(4), pp. 770-776, (2005). 






\end{thebibliography}
\end{document}